\magnification =1200
\vsize = 23.6 truecm\hsize =16 truecm

\def\spose#1{\hbox to 0pt{#1\hss}}
\def\Libra{\spose {--} {\cal L}}


\centerline{\it Contribution  aux
Journ\'ees Relativistes 1979,  ed. I. Moret-Bailly \& C. Latremoli\`ere,}
\centerline {\it ``Anneau d'accr\'etion sur les trous noirs''
pp 166-182 (Facult\'e des Sciences, Angers, 1979).}

\vskip 1 cm

\centerline{\bf GENERALISED MASS VARIATION FORMULA}
\centerline{ \bf FOR A STATIONARY AXISYMMETRIC STAR OR}
\centerline{\bf BLACK HOLE WITH SURROUNDING ACCRETION DISC}

\bigskip
\centerline{\bf Brandon Carter}
\centerline{G.A.R., Observatoire de Paris, 92 Meudon.}

\vskip 1 cm

\parindent = 0 cm

 {\bf Abstract} {\it The general relativistic mass-energy variation 
formula for axisymmetric equilibrium states of a selfgravitating 
system is developed in the particular case for which the relevant
matter consists of a perfectly conducting multiconstituent fluid 
(or superfluid) with  stationary circular or convective motion.}

\vskip 1 cm

\parindent= 0 cm
 {\bf 1. Introduction}
\medskip\parindent=1 cm

This communication presents an extension of the study$^1$ of mass
variation formulae for stationary axisymmetric systems -- relativistic
stars and black holes with surrounding material discs -- that is
contained in my contribution (chapter 6) to the Einstein Centenary
Survey: General Relativity, ed S.W. Hawking and W. Israel
(Cambridge U.P. 1979); all numbered references here refer to equations
presented therein.

The present work starts from the general formula
$$\delta M ={1\over 8\pi}\delta\int G_a^{\ b} k^a\, {\rm d}\Sigma_b -
{1\over 8\pi}\int G^{cd}h_{cd} k^a\, {\rm d}\Sigma_a +\Omega^{_{\rm H}}\, 
\delta J_{_{\rm H}} +{\kappa\over 8\pi}\, \delta{\cal A} \eqno{(1)} $$
(equation (6.342)$^1$ ) for the variation of the total asymptotically
defined mass $M$ of the system between two neighbouring stationary
axisymmetric equilibrium states, where $\Omega^{_{\rm H}}$, 
$J_{_{\rm H}}$, $\kappa$, ${\cal A}$ are the angular velocity, 
angular momentum, decay parameter (``surface gravity'') and 
surface area of the central black hole -- if it is present -- while 
$G_{ab}$ is the Einstein tensor, $h_{cd}$ the variation of the 
metric tensor, $k^a$ the stationary Killing vector, and the integration
is taken over a spacelike hypersurface $\Sigma$ extending out to the
asymptotically flat region with inner boundary on the black hole
if it is present. 

 \bigskip\parindent = 0 cm
{\bf 2. Electromagnetic and material contributions}
\medskip\parindent = 1 cm
 
We now introduce the energy momentum tensor $T_{ab}$ and the
corresponding momentum vector ${\mit \Pi}^a=-T^a_{\ b} k^b$ (decomposed 
into electromagnetic and purely material parts, labelled by suffices
$_{\rm F}$ and $_{\rm M}$ ) as follows. When we substitute from the
Einstein equations into the fundamental kinematic mass variation
formula (1) we obtain
$$ \delta M -\Omega^{_{\rm H}}\, \delta J_{_{\rm H}} -
{\kappa\over 8\pi}\, \delta{\cal A} = \delta\int T^a_{\ b} k^b\,
{\rm d}\Sigma_a -{_1\over^2}\int T^b_{\ c} h^c_{\ b} k^a\, {\rm d}
\Sigma_a \eqno{(2)}$$
 i.e.
$$ \delta M -\Omega^{_{\rm H}}\, \delta J_{_{\rm H}} -
{\kappa\over 8\pi}\, \delta{\cal A} +\delta\int{\mit \Pi}_{_{\rm F}}^{\ a}
\{k\}\, {\rm d}\Sigma_a+{_1\over^2}\int T^b_{_{\rm F} c} h^c_{\ b} k^a\, 
{\rm d}\Sigma_a $$
$$ =-\delta\int{\mit \Pi}_{_{\rm M}}^{\ a}
\{k\}\, {\rm d}\Sigma_a-{_1\over^2}\int T^b_{_{\rm M} c} h^c_{\ b} k^a\, 
{\rm d}\Sigma_a \eqno{(3)}$$

To evaluate the electromagnetic contribution we first use the
analogue of (6.306) for the momentum flux of a stationary field,
namely
$$ -{\mit \Pi}_{_{\rm F}}^{\ a} =\Lambda_{_{\rm F}} k^a+\big( k^b A_b\big)
 j^a+\nabla_{\! b}\big( A_c k^c D_{_{\rm F}}^{\ ba}\big) \eqno{(4)}$$
and then work out
$$ |g|^{-1/2}\delta\big(|g|^{1/2}\Lambda_{_{\rm F}}k^a\big)={_1\over^2}
\big(T_{_{\rm F}\ c}^{\ b} h^c_{\ b}-D_{_{\rm F}}^{\ bc}
\delta F_{bc}\big) k^a  $$
$$ \hskip 3 cm=\big({_1\over^2}T_{_{\rm F}}^{\ bc} h_{bc}
-j^b\delta A_b\big) k^a +2\nabla_{\! b}\big(D_{_{\rm F}}^{\ c[b}
\delta A_c k^{a]}\big)\, , \eqno{(5)}$$
where the electromagnetic action density 
and displacement tensor are given -- by (6.193) and (6.194)$^1$ -- 
as $\Lambda_{_{\rm F}}=(1/16\pi)F_{ab}F^{ba}$ and
$D_{_{\rm F}}^{\ ab}=(1/4\pi)F^{ab}$.

With the aid of the Green theorem (and using the asymptotic flatness
conditions to eliminate a surface contribution at infinity) we are
thus able to obtain
$$ \delta\int_\Sigma {\mit \Pi}_{_{\rm F}}^{\, a}\,{\rm d}\Sigma_a
-{_1\over^2} \int_\Sigma T_{_{\rm F}\ c}^{\ b} h^c_{\ b}k^a\,
{\rm d}\Sigma_a =\int_\Sigma k^b A_b\,\delta\big( j^a\,
{\rm d}\Sigma_a\big) +{_1\over^2}\int_\Sigma k^{[b} j^{a]}
\big(\delta A_b\big)\, {\rm d}\Sigma_a $$
$$\hskip 3 cm -{_1\over^2} \oint_{\cal S}{_1\over^2} k^c A_c 
D_{_{\rm F}}^{\ ba}\,{\rm d}{\cal S}_{ab}-\oint_{\cal S} 
D_{_{\rm F}}^{\ c[b} \big(\delta A_c\big) k^{a]}\, {\rm d}{\cal S}_{ab}
\eqno{(6)}$$

\bigskip\parindent = 0 cm
{\bf 3. Perfectly conducting fluid}
\medskip\parindent = 1 cm

          The result so far does not depend on the particular field 
equations satisfied by any matter that is present. In order to 
show the relationship with the first law of ordinary thermodynamics 
we shall now consider the explicit form of the right hand side of (3)
in a particularly ideal case namely that of a {\it perfectly conducting
multiconstituent fluid medium}. Previous descriptions of the first
law by Bardeen 1973$^2$, Bardeen, Carter and Hawking 1973$^3$ have been
restricted to the more specialised case of a {\it non-conducting} medium, 
electromagnetic effects being taken into account in the more 
extended treatment given by Carter 1973b$^4$. 

            The more general treatment presented here is based on the 
relativistic theory of a multiconstituent perfect fluid obtained as the 
natural extension of the two constituent case discussed by Carter 
1976$^5$. According to this theory (which includes the Landau-London 
two-fluid model for superfluidity as a special case in the 
non-relativistic limit) the material equations of motion are derivable 
from a Lagrangian scalar $\Lambda_{_{\rm M}}$ which is a covariant
function of a set of conserved current vectors $\vec{n}_{_X}$ 
representing the fluxes of a set of independent constituents labelled by 
an index $_X$, one of the constituents being usually (except in 
zero-temperature models) the {\it entropy}. (Thus in a typical application  
$_X$ might take the values $ _{\rm I}$, $ _{\rm I\! I}$, 
$ _{\rm I\! I\! I}$ with $\vec{n}_{_{\rm I}}$ and 
$\vec{n}_{_{\rm I\! I}}$ representing fluxes of positively charged 
ions and negatively charged electrons respectively, while  
$\vec{n}_{_{\rm I\! I\! I}}$ is identified with an entropy flux
vector $\vec{\bf s}$.

          Under the influence of arbitrary infinitesimal variations of the 
fluxes $\vec{n}$ and of the metric ${\underline  g}$ the function 
$\Lambda_{_{\rm M}}$ will have an infinitesimal variation of the form
$$\delta \Lambda_{_{\rm M}}=\mu^{_X}_{\, a}\delta n_{_X}^{\, a}
+{_1\over^2} \mu^{_X}_{\, a} n_{_X}^{\, b} h^a_{\, b} \, .\eqno{(7)}$$
The covectors $\underline \mu^{_X}$ so defined may appropriately be described
as the {\it chemical potential one-forms} of the corresponding conjugate
constituents. (In particular the covector conjugate to the entropy
flux vector $\vec s$ may appropriately be denoted by $\underline \Theta$ and
interpreted as a {\it temperature form}.) Their constituent indices
are written upstairs in view of the fact that they transform
{\it contravariantly} with respect to the fluxes under a change of
{\it chemical basis} in the vector space of constituents.

          When the medium interacts with an electromagnetic field the
{\it total} Lagrangian will be taken to have the simple additive form
$$ \Lambda = \Lambda_{_{\rm M}}+A_a j^a+\Lambda_{_{\rm F}} \eqno{(8)}$$
where $\Lambda_{_{\rm F}}$ is the electromagnetic Lagrangian
(given by (6.193)$^1$) and where the electromagnetic current $j^a$
appearing in the interaction term is given by
$$ \vec j= e^{_X} \vec n_{_X} \eqno{(9)}$$
where $e^{_X}$ represents the mean {\it electric charge per particle}
of the $_X$th constituent. Like the chemical potentials, the charges
behave as components of a contravariant vector in the chemical
constituent space. (It would in fact be easy to work with a more
general Lagrangian involving a more intimate relationship between 
the fluxes $\vec n_{_X}$ and the field $\underline F$. The restriction
to the simple additive form (8) -- which means that polarisability
effects are excluded -- has been made merely to facilitate the
interpretation of the formalism.) Variation of the total
Lagrangian now gives
$$ \delta \Lambda \equiv \pi^{_X}_{\, a} \delta n_{_X}^{\, a}+
(j^a-j_{_{\rm F}}^{\,a})\delta A_a - \nabla_{\! c} \big( 
D_{_{\rm F}}^{\, cb}\delta A_b\big)\hskip 2 cm $$
$$\hskip 3  cm +{_1\over^2}\Big( n_{_X}^{\,a}\pi^{_X}_{\, b} -
D_{_{\rm F}}^{\, ca}\nabla_{\! b} A_c+\nabla_{\! c}\big(D_{_{\rm F}}^{\ ca} 
A_b\big)+(j_{_{\rm F}}^{\, a}- j^a)A_b\Big) h_a^{\ b} \eqno{(10)}$$
where we have used the abbreviation
$$ j_{_{\rm F}}^{\, a} \equiv \nabla_{\! b} D_{_{\rm F}}^{\, ab}
\eqno{(11)}$$
in accordance with the scheme introduced in section 6.4$^1$, and
where we introduce the notation
$$\underline\pi^{_X}=\underline\mu^{_X} + e^{_X}\underline  A \eqno{(12)}$$
for what will play the role of an effective {\it four-momentum per 
particle} for the corresponding constituent. If we required the
volume integral of $\Lambda$ to be invariant under {\it arbitrary}
variations of the fields $\underline A$ and $\vec n_{_X}$ we would obtain
not only the correct equations of motion
$$\vec j_{_{\rm F}}= \vec j \eqno{(13)}$$
for the electromagnetic field, but also the too restrictive requirement
that the momenta $\underline\pi^{_X}$ vanish altogether. To obtain
reasonable equations of motion for the chemical constituents one
must demand invariance not under arbitrary current variations but only 
under variations of the restricted class that arise from {\it 
infinitesimal displacements of the world lines} of the idealised
particles, or directly from variations of $g_{ab}$. Now it can easily
be seen (e.g using the formulae given by Carter 1973a$^6$) that an
infinitesimal displacement $\vec \xi_{_X}$ of the world lines of a
conserved flux of particles (acting in conjunction with a variation
$\underline h$ of the metric) induces a corresponding variation
$$ \delta n^a = -{_1\over^2} n^a h^c_{\, c} - n^a \nabla_{\! b} \xi^b
+ n^b\nabla_{\! b} \xi^a - \xi^b\nabla_{\! b} n^a \eqno{(14)}$$
in the corresponding current vector $\vec n$. 

       The equations (7) to (13) were {\it chemically covariant} 
(i.e. unaffected by changes in the chemical basis of the space of 
constituents) in {\it form} although not necessarily in content (since 
the {\it specific} representation of the Lagrangian will in general 
depend on the chemical basis). However the choice of a particular set of 
chemical constituents implies the choice of a {\it preferred basis} in 
the vector space of constituents, so that the system of equations that 
results will not have even the appearance of chemical covariance. It is 
also to be remarked that although the identities derived in section 
6.4.1$^1$ (which depended only on the covariance of the Lagrangian under
general co-ordinate transformations of the space-time manifold as
opposed to the chemical constituent vector space) will still be
formally valid for the material lagrangian $ \Lambda_{_{\rm M}}$  -- 
indeed (6.135)$^1$ has been implicitely used in the derivation of the
form (7) -- and for the total Lagrangian $\Lambda$, the quantities
involved cannot be given the same direct interpretation (as representing
energy-momentum etc. ...) for a {\it restrained} variational theory as
in the free field theory for which they were originally intended. In the
present theory the appropriate energy-momentum tensor cannot be obtained
directly from the analogue of (6.138)$^1$ but can be read out from the
expression
$$ |g|^{-1/2}\delta\big( |g|^{1/2}\Lambda \big)\equiv T^{ab} h_{ab}+
(j^a -j_{_{\rm F}}^{\, a})\delta A_a - f^{_X}_{\, a}\xi_{_X}^{\, a}
+\nabla_{\! a}\Big( 2\pi^{_X}_{\, b}n_{_X}^{[a}
\xi_{_X}^{\, b]}- A_b D_{_{\rm F}}^{\, ab}\Big) \eqno{(15)}$$
where the $\vec \xi_{_X}$ (which do {\it not} behave as components
of a well behaved vector under chemical basis transformations) are a
set of arbitrary displacement vector fields acting on the world lines
of the corresponding fluxes, and where the fields $\underline f^{_X}$ 
(to which the same remark applies) are to be interpreted as generalised
{\it 4-force densities} acting on the particles of the corresponding
constituents. One finds that the force covectors have the explicit
form 
$$ \underline f^{_X}\equiv \vec n_{_{|X|}}\!\cdot\partial\underline
\pi^{_X}+\underline\pi^{_X}\nabla\cdot\vec n_{_{|X|}} \eqno{(16)}$$
where the inclusion of the chemical indices within {\it straight bars}
is introduced to denote suspension of the summation convention.

Each of the variational field equations
$$ \underline f^{_X}=0           \eqno{(16)}$$
can be decomposed by first contracting with the corresponding flux
$\vec{n}^{_X}$, to give {\it first} the conservation law
$$ \nabla\cdot\vec  n_{_X}=0 \eqno{18}$$
for each constituent, and {\it then} the canonical equation of motion
$$ \vec n_{_{|X|}}\!\cdot \partial\underline \pi^{_X} =0\eqno{(19)}$$
which determines the acceleration of the corresponding world lines. 
This last equation may be decomposed into a material and an
electromagnetic part in the form
$$ \vec n_{_{|X|}}\!\cdot \partial\underline \mu^{_X}
=\underline F\!\cdot\vec j_{^X}   \eqno{(20)}$$
where
$$ \vec j^{_X}=e^{_X}\vec n_{_{|X|}} \eqno{(21)}$$
is the electric current density associated with the $_X$th constituent.
It is to be remarked that the condition (19) is precisely what is
required for the flux of generalised vorticity
$$ \underline w^{_X}= \partial\underline \pi^{_X} \eqno{(22)}$$
to be conserved along the flow lines in the strong sense discussed
in section 6.4.5$^1$, meaning that
$$(\alpha \vec{n}_{_{|X|}}\!){\Libra} \underline w^{_X}=0\eqno{(23)}$$
for an {\it arbitrary} scalar field $\alpha$, which implies that the
{\it flux} $\underline w^{_X}$ over any 2-surface, which by Stokes
theorem is equal to the {\it circulation}
$$ {\cal C}^{_X }=\oint_c\underline \pi^{_X}\!\cdot{\rm d}\vec x
\eqno{(24)}$$
around its boundary circuit $c$ say, is conserved in the sense of
being unchanged when the circuit (or the original 2-surface enclosed
by it) is dragged at an arbitrary rate along the flow lines.

Unlike the field equations themselves, the expression for the {\it energy
momentum} that is read out from (15) has a chemically covariant form.
After subtracting out the Maxwellian contribution in accordance with
the decomposition (6.299)$^1$ we are left with a {\it material}
contribution to the energy-momentum given by
$$ T_{_{\rm M}\ b}^{\ a}=n_{_X}^{\, a} \mu^{_X}_{\, b}
-\Psi_{_{\rm M}} g^a_{\,b} \eqno{(25)}$$
where $\Psi_{_{\rm M}}$ is a {\it thermodynamic potential density}
that is related to the material Lagrangian by
$$\Psi_{_{\rm M}}=n_{_X}^{\, a}\mu^{_X}_{\, a}-\Lambda_{_{\rm M}}
\ .\eqno{(26)}$$
{\it Provided} that the electromagnetic field equations (13) are
satisfied, the expression (6.299)$^1$ of the total metric energy
momentum tensor as the sum of metric contributions $T_{_{\rm F}}^{\ ab}$ 
and $T_{_{\rm M}}^{\ ab}$ can be replaced by an expression in terms of 
canonical contributions in the form 
$$ T^a_{\ b}={\cal T}_{_{\rm MC}\, b}^{\ a}+{\cal T}_{_{\rm F}\ b}^{\, a}
+\nabla_{\!c}\big(D_{_{\rm F}}^{\ ca} A_b\big) \eqno{(27)}$$
where ${\cal T}_{_{\rm F}\ b}^{\ a}$ are components of the canonical
electromagnetic energy momentum tensor introduced in (6.145)$^1$, and
where the corresponding {\it canonial} material energy momentum tensor
is defined by
$${\cal T}_{_{\rm MC}\, b}^{\ a}=n_{_X}^{\, a} \pi^{_X}_{\, b}
-\Psi_{_{\rm M}} g^a_{\,b}\ . \eqno{(28)}$$
Subject to the field equations, the {\it total} energy momentum tensor
will of course be preserved, i.e.
$$ \nabla_{\!b}T^b_{\ a}=0 \eqno{(29)}$$
and hence the separate parts will satisfy
$$\nabla_{\! b} T_{_{\rm M}\, a}^{\ b}= F_{ab}j^b=-
\nabla_{\! b} T_{_{\rm F}\ a}^{\ b}\ ,\eqno{(30)}$$
$$\nabla_{\! b}{\cal T}_{_{\rm MC}\, a}^{\ b}=j^b \nabla_{\!a}A_b
= -\nabla_{\! b}{\cal T}_{_{\rm F}\ a}^{\ b}\ .\eqno{(31)}$$

For any vector field $\vec \xi$ we define associated canonical and
metric energy - momentum flux vectors analogous to those introduced
in section 6.4$^1$ by
$$ \vec{\cal P}\{\xi\}=\vec{\cal P}_{_{\rm MC}}\{\xi\}+
\vec{\cal P}_{_{\rm F}}\{\xi\}\ , \hskip 1 cm
{\cal P}_{_{\rm MC}}^{\ a}\{\xi\}= -{\cal T}_{_{\rm MC}\, b}^{\ a}
\,\xi^b\ .\eqno{(32)}$$
$$ \vec{\mit \Pi}\{\xi\}=\vec{\mit \Pi}_{_{\rm M}}\{\xi\}+
\vec{\mit \Pi}_{_{\rm F}}\{\xi\}\ , \hskip 1 cm
{\mit \Pi}_{_{\rm M}}^{\ a}\{\xi\}= -T_{_{\rm M}\ b}^{\ a}
\,\xi^b\ ,\eqno{(33)}$$
the total being related by
$${\cal P}^{a}\{\xi\}={\mit \Pi}^{a}\{\xi\}+\nabla_{\! b}\big(
D_{_{\rm F}}^{\ ba} A_c\xi^c\big)\ .\eqno{(34)}$$
It is obvious from (29) that the {\it total} metric energy - momentum
fluxes will be conserved whenever $\vec\xi$ is a Killing vector, and
hence that the same applies to the canonical energy - momentum flux,
i.e.
$$ \hskip 2 cm \vec\xi{\Libra}\,\underline g=0 \hskip  1.5 cm\Rightarrow 
\hskip  1.5 cm \bigg\{ {\nabla\cdot\vec{\mit \Pi}\{\xi\}=0\atop \
\nabla\cdot\vec{\cal P}\{\xi\}=0\, . } \hskip 3 cm {(35)\atop (36)}$$
However the {\it canonical} energy - momentum flux also has the property
that is {\it material part} will be conserved {\it separately} whenever
the fields are {\it themselves invariant}. Clearly the equation of motion
(19) implies (by the Cartan formula (6.20)$^1$) that we shall have a
generalised Bernoulli theorem to the effect that
$$\hskip 2 cm \vec\xi{\Libra}\underline\pi^{_X}=0 \hskip 1.6 cm \Rightarrow
\hskip 1.6 cm \vec n_{_{|X|}}\!\cdot\nabla(\vec\xi\cdot\underline\pi^{_X})
=0\ ,\hskip 3 cm \eqno{(37)}$$
i.e. that $\vec \xi\cdot\underline\pi^{_X}$ is a {\it constant of the
motion}, and hence, using (18) we see that 
$$ {  \vec\xi{\Libra}\underline \pi^{_X}= 0\ ,\hskip 1 cm  \vec\xi{\Libra}
\Psi_{_{\rm M}}= 0 \atop \vec\xi{\Libra}\,\underline g=0}
\biggr\}
\hskip 1 cm\Rightarrow \hskip 1 cm \nabla\cdot \vec
{\cal P}_{_{\rm MC}}\{\xi\}=0\ . \eqno{(38)}$$
This is to be compared with the corresponding result (6.157)$^1$ for
the electromagnetic component, which tells us that the same result would
follow from purely electromagnetic invariance, i.e.
$$ {  \vec\xi{\Libra}\underline A= 0 \atop \vec\xi{\Libra}\,
\underline g=0}\biggr\}\hskip 1 cm\Rightarrow \hskip 1 cm 
\nabla\cdot \vec{\cal P}_{_{\rm MC}}\{\xi\}=0\ . \eqno{(39)}$$
It may be noticed that across any {\it invariant} hypersurface, i.e.
one whose surface element satisfies
$$  \vec\xi\cdot {\rm d}\underline\Sigma=0 \eqno{(40)}$$
the canonical material energy - momentum flux is interpretable as the
{\it total} rate of transport of the Bernoulli quantity, i.e.
$$ {\cal P}_{_{\rm MC}}^{\ a}\{\xi\}\, {\rm d}\Sigma_a= (\vec\xi\cdot
\underline\pi^{_X}) {\rm d} N_{_X} \eqno{(41)}$$
where ${\rm d} N_{_X}$ denotes the number flux of the $_X$th
constituent across ${\rm d}\underline\Sigma$ , i.e.      
$${\rm d} N_{_X}= -\vec n_{_X}\!\cdot{\rm d}\underline\Sigma \ .\eqno{(42)}$$

 \bigskip\parindent = 0 cm
{\bf 4. Global energy and angular momentum}
\medskip\parindent = 1 cm

If we now consider the particular case of a stationary and/or
axisymmetric system, in which there will be a conserved energy and/or
angular momentum per particle given by
$$   {\cal E}^{_X}= -\vec k\cdot \underline\pi^{_X} \eqno{(43)}$$
$$   {\cal J}^{_X}= \vec m\cdot \underline\pi^{_X} \eqno{(44)}$$
then we shall have
$$ {\cal P}_{_{\rm MC}}^{\ a}\{k\}\, {\rm d}\Sigma_a = -{\cal E}^{_X}
{\rm d}N_{_X}=\sum_{_X} \vec{\cal E}_{_{\rm CM}}^{\ _X}\cdot
{\rm d}\underline\Sigma \eqno{(45)}$$
$$ {\cal P}_{_{\rm MC}}^{\ a}\{m\}\, {\rm d}\Sigma_a = {\cal J}^{_X}
{\rm d}N_{_X}= -\sum_{_X} \vec{\cal J}_{_{\rm CM}}^{\ _X}\cdot
{\rm d}\underline\Sigma \eqno{(46)}$$
across any respectively stationary or axisymmetric hypersurface, and
in particular across a black hole horizon, where we define the fluxes
$$\vec{\cal E}_{_{\rm MC}}^{\ _X}={\cal E}^{_X}\vec n_{_{|X|}} \eqno{(47)}$$
$$\vec{\cal J}_{_{\rm MC}}^{\ _X}={\cal J}^{_X}\vec n_{_{|X|}} \eqno{(48)}$$
which evidently satisfy the separate conservation laws
$$\nabla\cdot\vec{\cal E}_{_{\rm MC}}^{\ _X}=0\eqno{(49)}$$
$$\nabla\cdot\vec{\cal J}_{_{\rm MC}}^{\ _X}=0\ .\eqno{(50)}$$
In the case of angular momentum, the total exterior source contribution
given by (6.322)$^1$ may be expressed by
$$ J_{_{\rm MC}}=\sum_{_X}\int_\Sigma\vec{\cal J}_{_{\rm MC}}^{\ _X}
\cdot{\rm d}\underline\Sigma\eqno{(51)}$$
for any axisymmetric hypersurface $\Sigma$, but the fact that $\Sigma$
could not be invariant under $\vec k$ prevents us from obtaining an
equaly simple formula for $M_{_{\rm MC}}$.

We now proceed from (6) as in Section 6.6.1 using the strict stationary
boundary conditions (6.314)$^1$ to evaluate the surface integrals on
the horizon in the form
$$\delta\oint_{\cal S}{_1\over^2} k^c A_c D_{_{\rm F}}^{\, ba}\,
{\rm d}{\cal S} -\oint_{\cal S}D_{_{\rm F}}^{\, c[b}(\delta A_c)k^{a]}
\,{\rm d}{\cal S}_{ab}
=\Phi^{_{\rm H}}\delta Q_{_{\rm H}}+\Omega^{_{\rm H}}\delta J_{_{\rm FH}}
\eqno{(52)} $$
We next perform the analogous steps in the evaluation of the material
contributions, using the formulae of the previous section which enable
us to write firstly
$$ -{\mit \Pi}_{_{\rm M}}^{\ a}\{k\}=\big(\Lambda_{_{\rm M}}-n_{_X}^{\, b}
\mu^{_X}_{\, b}\big)k^a+ \big(k^b\mu^{_X}_{\, b}\big)n_{_X}^{\, a}
\eqno{(53)}$$
and
$$ |g|^{-1/2}\delta\big( |g|^{1/2}\Lambda_{_{\rm M}}\big)={_1\over^2}\big( 
T_{_{\rm M}\, c}^{\ b} h^c_{\, b}+n_{_X}^{\, b}\mu^{_X}_{\, b} h^c_{\, c}
\big)+\mu^{_X}_{\,b}\delta n_{_X}^{\,b}\ .\eqno{(54)}$$
Since we now only have source terms to deal with we proceed directly
(i.e. without recourse to Green's theorem)  using (6.399)$^1$ to obtain
$$-\delta \int_\Sigma {\mit \Pi}_{_{\rm M}}^{\,a}\,{\rm d}\Sigma_a-
{_1\over^2}\int_\Sigma T_{_{\rm M}\, c}^{\ b} h^c_{\, b} k^a\,
{\rm d}\Sigma_a=\int_\Sigma k^b \mu^{_X}_{\, b}\delta\big(n_{_X}^{\,a}\,
{\rm d}\Sigma_a\big)+{_1\over^2}\int_\Sigma k^{[b} n_{_X}^{\, a]}\big(
\delta \mu^{_X}_{\,b}\big){\rm d}\Sigma_a\eqno{(55)}$$
in which the terms can be seen to be closely analogous to the volume
integral terms in (6). Thus using the expression (12) for the {\it total
momenta} we can combine (6) and (55) so as to obtain (with the aid of
(52) and (6.320)$^1$)  the final result
$$\delta M-\Omega^{_{\rm H}}\delta J_{_{\rm BH}}
-\Phi^{_{\rm H}}\delta Q_{_{\rm H}}-
{\kappa\over 8\pi}\delta{\cal A}=\int_\Sigma \vec k\cdot\underline\pi^{_X}
\delta\big(\vec n_{_X}\!\cdot {\rm d}\underline\Sigma\big)+\int_\Sigma
\big(\delta\underline\pi^{_X}\big)\!\cdot\big(\vec k\wedge\vec n_{_X}\big)
\cdot {\rm d}\underline\Sigma\ \eqno{(56)}$$
(in which $J_{_{\rm BH}}=J_{_{\rm H}}+J_{_{\rm FH}}= J-J_{_{\rm MC}}$).

 \bigskip\parindent = 0 cm
{\bf 5. Interpretation in the generic (convective) case}
\medskip\parindent = 1 cm

The first term on the right hand side of (56) can be interpreted as
representing the {\it static} injection energy: it has the form
$$\int_\Sigma \vec k\cdot\underline\pi^{_X}
\delta\big(\vec n_{_X}\!\cdot {\rm d}\underline\Sigma\big)=\int
{\cal E}^{_X}\delta\big({\rm d} N_{_X}\big) \eqno{(57)}$$
where ${\rm d} N_{_X}$, as given by (42), is the flux of the $_X$th
constituent across ${\rm d}\underline\Sigma$, and where ${\cal E}^{_X}$
is the canonical {\it energy per particle}, as given by (43), which
satisfies the conservation law
$$ \vec n_{_{|X|}}\!\cdot\nabla {\cal E}^{_X}=0\eqno{(58)}$$
for each $_X$.

The form of the second term in (56) (which vanishes altogether in a
static situation, i.e. one with all current vectors parallel to $\vec k$)
may seem less familiar, but it is in fact a generalisation of the
Faraday term $I\delta{\cal F}$ representing the energy needed to make a 
change $\delta {\cal F}$ in the magnetic flux ${\cal F}$ through a circuit
carrying a current $I$. The relationship can be made rather more apparent
if we decompose each flux $\vec n_{_X}$ into a {\it static} part plus a
spacelike part $\vec c_{_X}$ that is confined within the spacelike 
hypersurface $\Sigma$, in the form
$$ \vec n_{_X}= n_{_X}^{_{(0)}}\vec k+\vec c_{_X}\ ,\hskip 1.5 cm
\vec c_{_X}\!\cdot{\rm d}\underline\Sigma=0\eqno{(59)}$$
since we may then write
$$\int_\Sigma\big(\delta\underline\pi^{_X}\big)\!\cdot\big(\vec k
\wedge\vec n_{_X}\big)\cdot {\rm d}\underline\Sigma= -\int
\big(\vec c_{_X} \delta\underline\pi^{_X}\big)\big(\vec k\!\cdot
{\rm d}\underline \Sigma\big)\ .\eqno{(60)}$$

We are always at liberty to decompose the hypersurface element
${\rm d}\underline\Sigma$ in the form
$$ {\rm d}\underline\Sigma={\rm d}\vec x\!\cdot {\rm d}
\underline{\cal S}\eqno{(61)}$$
where ${\rm d}\vec x$ is an arbitrary infinitesimal displacement and
${\rm d}\underline{\cal S}$ is a normal 2 surface element transverse
to the displacement. If we take ${\rm d}\vec x$ to be locally parallel
to one of the fluxes, $\vec c_{_X}$ say, we shall have
$$ -\big(\vec c_{_{|X|}} \delta\underline\pi^{_X}\big)\big(\vec k\!\cdot
{\rm d}\underline \Sigma\big)=\big({\rm d}\vec x\!\cdot\delta
\underline\pi^{_X}\big) {\rm d} C_{_{|X|}}\eqno{(62)}$$
where ${\rm d} C_{_{|X|}}$ represents the current of the $_X$th constituent
through the element ${\rm d}\underline{\cal S}$ as defined by
$${\rm d} C_{_X}=\vec c_{_X}\!\cdot\big(\vec k\!\cdot {\rm d}
\underline{\cal S}\big)\ .\eqno{(63)}$$ 
Thus ${\rm d} C_{_X}$ represents the number of particles crossing the
element ${\rm d}\underline{\cal S}$ per unit of the {\it ignorable}
(not proper) group time $t$. In general one can envisage that the flux
lines in $\Sigma$ will have complicated ergodic properties. However
{\it if} their global structure is such that they form sufficiently
well behaved {\it simple closed circuits} then one can see from (62)
that we shall have 
$$\delta M-\Omega^{_{\rm H}}\delta J_{_{\rm BH}}
-\Phi^{_{\rm H}}\delta Q_{_{\rm H}}-
{\kappa\over 8\pi}\delta{\cal A}=\int{\cal E}^{_X}\delta
\big({\rm d} N_{_X}\big)+\int \big(\delta {\cal C}^{_X}\big)
{\rm d} C_{_X} \eqno{(64)}$$
where for each circuit the conserved circulation integral ${\cal C}^{_X}$,
which is a generalisation of the Faraday flux, is as defined by (24).

 \bigskip\parindent = 0 cm
{\bf 6. The circular (non - convective) case}
\medskip\parindent = 1 cm

Despite the restriction imposed by the non-ergodicity hypothesis, the
deceptively simple form (64) still represents quite a powerful
generalisation compared not only with the original variation law
of Bardeen {\it et al} (1973)$^{2,3}$, but even with the electromagnetic
extension of Carter (1973b)$^4$, since these earlier versions applied
only to flows with strictly (not just topologically) circular circuits.
To show how these earlier versions are obtained in the appropriate,
non - convective (and non - conducting) limit, we first introduce a
vector $\vec v_{_X}$ defined by
$$\vec c_{_X}= n_{_X}^{\, _{(0)}} \vec v_{_{|X|}} \eqno{(65)}$$
and representing the mean {\it displacement} of the particles of the
$_Xth$ constituent  per unit of the {\it ignorable} (not proper) group
time $t$. (If there were any particles in the immediate neighbourhood of 
the black hole, their 3-velocities as thus defined would evidently have
to tend towards the Damour$^7$ velocity $\vec v_{_{\rm H}}$ -- as defined
in section 6.3$^1$ -- in the limit on the horizon.) We see that in terms
of the corresponding unit flow vectors $\vec u_{_X}$, defined by
$$ \vec n_{_X}=n_{_X}  \vec u_{_{|X|}}\ ,\hskip 1.5 cm |\vec u_{_{|X|}}|^2
=-1 \eqno{(66)}$$
the 3-velocity 4-vectors will be given by
$$\vec k+\vec v_{_X}=\dot\tau_{_{|X|}}\vec u_{_X}\eqno{(67)}$$
where $\dot\tau_{_X}$ is the relevant time dilation (or ``mean redshift'')
factor. We can use these 3-velocities to regroup the terms on the
right hand side of (56) so as to obtain the variation law

$$\delta M-\Omega^{_{\rm H}}\delta J_{_{\rm BH}}-\Phi^{_{\rm H}}
\delta Q_{_{\rm H}}-{\kappa\over 8\pi}\delta{\cal A}=\int_\Sigma 
\pi^{_X}_{\,_{(0)}}\delta\big({\rm d} N_{_X}\big)-\int_\Sigma 
\vec v_{_X}^{\, b}\delta \big(\pi^{_X}_{_, b} n_{_X}^{\, a}\,{\rm d}
\Sigma_a\big) \eqno{(68)} $$
where
$$ \pi^{_X}_{\,_{(0)}}=-\big(\vec k+\vec v_{_{|X|}}\big)\cdot 
\underline \pi^{_X} \eqno{(69)}$$
which is valid quite generally (regardless of the topological
behaviour of the flow lines). 

If we now impose the restriction that
the flow lines be strictly {\it circular}, so that the 3-velocities
take the form 
$$\vec v_{_X}=\Omega_{_X} \vec m \eqno{(70)}$$
where $\Omega_{_X}$ are the corresponding {\it angular velocities} then
we shall have
$$\delta M-\Omega^{_{\rm H}}\delta J_{_{\rm BH}}
-\Phi^{_{\rm H}}\delta Q_{_{\rm H}}-
{\kappa\over 8\pi}\delta{\cal A}=\int_\Sigma\pi^{_X}_{\,_{(0)}}
\delta\big({\rm d} N_{_X}\big)+\int_\Sigma  \Omega_{_X} \delta
\big( {\rm d} J_{_{\rm MC}}^{\ _X}\big) \eqno{(71)}$$
with
$$ {\rm d} J_{_{\rm MC}}^{\ _X}=\vec{\cal J}_{_{\rm MC}}^{\ _X}\!
\cdot{\rm d}\Sigma\eqno{(72)}$$
where the constituent angular momentum fluxes are given by (48). The
quantity $\pi^{_X}_{\,_{(0)}}$, which is a generalisation of what
Bardeen$^2$ has called the zero angular momentum injection energy,
is given by
$$\pi^{_X}_{\,_{(0)}}={\cal E}^{_X}-\Omega_{_{|X|}} {\cal J}^{_X}
=\mu^{_X}_{_{(0)}}+ e^{_X}\Phi_{_{|X|(0)}}\eqno{(73)}$$
where the {\it effective} chemical potential $\mu^{_X}_{_{(0)}}$
is related to the corresponding local rest frame chemical potential
$\mu^{_X}$ by
$$\mu^{_X}_{_{(0)}}=-\big(\vec k+\vec v_{_{|X|}}\big)\!\cdot
\underline\mu^{_X}=\dot\tau_{_{|X|}}\mu^{_X}\eqno{(74)}$$
with
$$\mu^{_X}=-\vec u_{_{|X|}}\!\cdot\underline\mu^{_X} \eqno{(75)}$$
and where the corotating electric potential $\Phi_{_{X(0)}}$
is given by
$$\Phi_{_{X(0)}}=-\big(\vec k+\vec v_{_{|X|}}\big)\!\cdot
\underline A=\dot\tau_{_{|X|}}\Phi_{_X}\eqno{(76)}$$
with
$$\Phi_{_X}=-\vec u_{_{|X|}}\!\cdot\underline A\ . \eqno{(77)}$$

\vfill\eject
\parindent=0 cm

{\bf References.}

\medskip

1. B. Carter, ``The general theory of the mechanical, electromagnetic
and thermodynamical properties of black holes'', in  {\it General
Relativity: an Einstein centenary survey}, ed S.W. Hawking \&
W. Israel {\it pp} 295 - 369 (Cambridge U.P. 1979).
  
\smallskip

2. J.M. Bardeen,  ``Rapidly rotating stars, discs and black holes'',
in {\it Black Holes}, ed B. DeWitt \& C. DeWitt {\it pp} 215 - 289
(Gordon \& Breach, New York, 1973).

\smallskip

3. J. Bardeen, B. Carter, \& S.W. Hawking, ``The Four Laws of Black 
Hole Mechanics", {\it Commun. Math. Phys.} {\bf 31}, 
{\it pp}  161-170 (1973).

\smallskip

4. B. Carter, ``Black Hole Equilibrium States: II General Theory of 
Stationary Black Hole States", in {\it Black Holes} 
(proc. 1972 Les Houches Summer School), ed. B. De Witt \& C. DeWitt, 
{\it pp} 125-210 (Gordon \& Breach, New York, 1973).

\smallskip

5. B. Carter, ``Regular and Anomalous Heat Conduction: the Canonical 
Diffusion Equation in Relativistic Thermodynamics", in {\it
 Journ\'ees Relativistes 1976}, ed. M. Cahen, R. Debever, \& J. Geheniau,
{\it pp} 12-27 (Universit\'e Libre de Bruxelles, 1976).

\smallskip

6. B. Carter,  ``Elastic Perturbation Theory in General Relativity and 
a Variational Principle for a Rotating Solid Star",
{\it Commun. Math. Phys.} {\bf 30}, {\it pp} 261-286 (1973).

\smallskip

7. T. Damour, ``Black hole eddy currents'',
 {\it Phys. Rev.} {\bf D18}, {\it pp} 3598-3604 (1978).

\end